\documentclass[final,5p,times,twocolumn]{elsarticle}

\usepackage{amssymb}
\usepackage{amsthm}
\usepackage{amsmath}
\usepackage{graphicx}
\usepackage[load-configurations = abbreviations]{siunitx}
\usepackage{booktabs}
\usepackage{lineno}
\usepackage{tabulary}

\newcommand{\um}[1]{ \SI{#1}{\micro\meter} }
\newcommand{\nm}[1]{ \SI{#1}{\nano\meter} }
\newcommand{\mm}[1]{ \SI{#1}{\milli\meter} }
\newcommand{\cm}[1]{ \SI{#1}{\centi\meter} }
\newcommand{\cmmc}[1]{\SI{#1}{\per\cubic\centi\metre}}
\newcommand{\fs}[1]{\SI{#1}{\femto\second}}
\newcommand{\GeV}[1]{\SI{#1}{\giga\electronvolt}}
\newcommand{\MeV}[1]{\SI{#1}{\mega\electronvolt}}
\newcommand{\pC}[1]{\SI{#1}{\pico\coulomb}}
\newcommand{\GVm}[1]{\SI{#1}{\giga\volt\per\meter}}

\journal{Nuclear Instruments and Methods in Physics Research A }

\begin{document}

\begin{frontmatter}



\title{Plasma boosted electron beams for driving Free Electron Lasers}


\author[a,f]{A.R. Rossi} 
\author[b]{V. Petrillo} 
\author[a]{A. Bacci} 
\author[c]{E. Chiadroni}
\author[c,d]{A. Cianchi}
\author[c]{M. Ferrario}
\author[c]{A. Giribono}
\author[c]{A. Marocchino}
\author[b]{M. Rossetti Conti}
\author[a]{L. Serafini} 
\author[c]{C. Vaccarezza}

\address[a]{INFN - MI, via Celoria 16, 20133 Milan, Italy}
\address[b]{University of Milan, Physics Department, via Celoria 16, 20133 Milan, Italy}
\address[c]{INFN - LNF, v.le E. Fermi, 00044 Frascati, Italy}
\address[d]{``Tor Vergata'' University, Physics Department, via della Ricerca Scientifica 1, 00133 Rome, Italy}
\address[f]{Corresponding author: andrea.rossi@mi.infn.it}

\begin{abstract}
In this paper, we report results of simulations, in the framework of both EuPRAXIA \cite{Walk2017} and EuPRAXIA@SPARC\_LAB \cite{Ferr2017} projects, aimed at delivering a high brightness electron bunch for driving a Free Electron Laser (FEL) by employing a plasma post acceleration scheme. The boosting plasma wave is driven by a tens of \SI{}{\tera\watt} class laser and doubles the energy of an externally injected beam up to \GeV{1}. The injected bunch is simulated starting from a photoinjector, matched to plasma, boosted and finally matched to an undulator, where its ability to produce FEL radiation is verified to yield $O(\num{e11})$ photons per shot at \nm{2.7}.     
\end{abstract}

\begin{keyword}
Laser Wakefield Acceleration \sep Electron Beam Dynamics \sep Free Electron Laser
\end{keyword}

\end{frontmatter}


\section{Introduction}

Laser WakeField Acceleration (LWFA) \cite{Taji1979} is considered one of the most promising candidate for future compact electron accelerators. It is widely believed that driving a Free Electron Laser (FEL) represents a significant goal to demonstrate the possibility to employ plasma based beams in most applications, spanning from industry to high energy physics frontiers.    

To date, many laboratories worldwide are investigating its potentialities \cite{plac}, using and testing different internal injection techniques, i.e. producing the electron beam by injecting background plasma electrons into the accelerating portion of a plasma wave. These settings allow for very compact accelerators (i.e. up to few tens of \cm{} in length) and output beams with energies up to few \GeV{} and charges from few to few hundreds of \pC{}. On the other hand, since control over injection remains quite flimsy, produced bunches display a number of unappealing features such as relatively high energy spreads and high divergences that induce beam emittance degradation in subsequent transport \cite{Floe2003,Anti2012,Migl2013}. Moreover, they may also suffer from unacceptable jitters in final energy, charge and pointing. Nevertheless, in the FEL framework, there exist the possibility to perform a chromatic matching \cite{Loul2015} to take advantage of the typical structure of internally injected longitudinal phase space beams and a recent project, named COXINELL \cite{Coup2016}, aims at demonstrating experimental feasibility. 

Schemes allowing to gain a complete control over the injected charge relay on an external source of electrons, generally a conventional accelerator \cite{ExIn}, and are designated with the name of external injection schemes or plasma boosters. 

In this paper we report numerical results showing energy doubling of an externally injected beam, up to a final energy of \GeV{1}, with preservation of beam quality around the peak of current. These simulations are part of start to end simulations aimed at proposing a possible LWFA external injection scheme for EuPRAXIA \cite{Walk2017} and EuPRAXIA@SPARC\_LAB \cite{Ferr2017} projects. Both projects have the goal of demonstrating the possibility to derive an FEL light sources for users, employing plasma based acceleration. In this design stage, there are no \emph{a priori} restrictions on acceleration techniques and methodology. While EuPRAXIA is an European Commission founded design study, EuPREAXIA@SPARC\_LAB is an INFN project for fostering the candidacy of INFN Frascati Labs to host EuPRAXIA, should it be approved.

The paper is structured as follows: in section \ref{requirements} performance requirements for EuPRAXIA beams and FEL are briefly summarized, in section \ref{bline} we briefly describe the beamline from photoinjector to undulator, with a focus on the plasma module. Section \ref{simulations} shows results of our simulations in plasma and within the undulator. Finally, in section \ref{conclusions}, we draw some conclusions.

\section{Beam requirements and FEL performances} \label{requirements}

Preliminary beam requirements for EuPRAXIA are reported in the deliverable D 1.2 \cite{EuDel}. LWFA with external injection from a RF linac is covered by Case 2, where an input beam with energy $\sim$ \MeV{150} and a current $\sim$ \SI{0.5}{\kilo\ampere} is assumed. As pointed out in the document, this setup would require a further beam compression for attaining the few \SI{}{\kilo\ampere} peak currents required for FEL operation. Indeed, beam dynamics considerations advise against relatively low energy beams with high currents, if top beam quality is required. The EuPRAXIA Plasma Wakefield Acceleration option (Case 4) foresees the presence of a \MeV{500} linac, allowing for a higher energy injection for LWFA, without requiring a subsequent magnetic compressor.    

The choice of FEL configuration and radiation scheme is based on user-defined requirements for properties of the output FEL pulses, i.e., radiation wavelength, peak power, polarization and required average repetition rate. One interesting spectral region of operation is the water window \SIrange{2.5}{4.0}{\nano\meter}, where users typically need $\sim$ \numrange{e12}{e13} photons/pulse in a 0.1\% bandwidth. Such radiation allows to study materials and biological tissues below the threshold of carbon and oxygen absorption.

\section{Beam line description} \label{bline}
In this section, we briefly describe the beamline from beam generation at photocathode to the undulator. Only the part reporting the design of the plasma booster module will include details, whereas other component are only summarized and relevant references will be indicated.

\subsection{Injector and linac}
A \pC{30} bunch is generated from a photoinjector and accelerated up to \MeV{100} in the injector \cite{Giri2017}, consisting in a \num{1.6} cell, S-band radio frequency (RF) gun and three RF traveling wave (TW), S-band accelerating cavities, i.e. a SPARC-like \cite{Ales2003} photoinjector. The gun and the first two TW cavities are surrounded by coils in order to perform emittance compensation \cite{emit_comp} and velocity bunching \cite{v_bunching}. The final TW cavity is operated slightly off crest to reduce energy spread.

The subsequent linac is composed by 32 X-band, \cm{50} long, TW structures \cite{Vacc2017} and is operated in order to increase beam energy up to a value of about \MeV{500} and reduce energy spread as much as possible, preserving beam emittance. After acceleration, the beam is matched into plasma by use of permanent magnets quadrupole. Nominal beam properties at plasma entrance are reported in Table \ref{linac_beam}.

\begin{table}[ht!]
   \centering
   \begin{tabular}{||c|c||} 
   \hline
   \hline 
    Charge (\SI{}{\pico\coulomb}) & \num{30} \\ \hline
    $\sigma_{x,y}$ (\um{}) & \num{0.7}, \num{1.2} \\ \hline
    $L_{\rm FWHM}$ (\um{}) & \num{3.2} \\ \hline
    $\varepsilon_{x,y}$ (mm-mrad) & \num{0.4}, \num{0.5} \\ \hline
    Energy (\SI{}{\mega\electronvolt}) & \num{536.5} \\ \hline
    $\Delta E/E$ & \num{6e-4} \\ \hline
    $I_{\rm peak}$ (\SI{}{\kilo\ampere}) & 2 \\  
   \hline
   \hline
   \end{tabular}
   \caption{Input bunch parameters. Emittance values are normalized r.m.s. values.}
   \label{linac_beam}
\end{table}

\subsection{Plasma module: plasma density and plasma wave regime}\label{plasma}

Established scaling laws for plasma acceleration \cite{Esar2009} suggest that a good choice for plasma wave regime in a booster is the quasi linear regime, arising when the laser strength parameter has a value $a_0 \approx 1$, together with an unperturbed plasma density $n_0 \approx$ \num{e17}\cmmc{}. An evaluation of the advantages related to this parameters choice, both from an electron beam dynamics and laser dynamics standpoint, was presented in \cite{Ross2014}. For chosen parameters values, the expected accelerating gradients have an upper theoretical limit of \GVm{30}; however, 3D effects and unavoidable non-ideal experimental set-ups, lead to a safer evaluation around \GVm{10}, so target electron energy of \GeV{1} should be achieved in an acceleration length around \cm{5}, which is short enough to avoid both dephasing and pump depletion effects \cite{Esar2009}, which become significant in length scales close to \SI{1}{\meter}. Energy spread growth may still constitute a problem because beams compressed by velocity bunching display a small FWHM length (few \um{}s) but usually also present a long tail in the trailing area. A correct injection in the plasma wave (i.e. the current peak resting on the longitudinal field peak) results in the tail being subject both to a defocusing force and a large accelerating gradient variation, leading to overall rms energy spread increase and, possibly, to charge loss and/or halo formation. However, since the tail usually comprises a small amount of total charge and has low slice current, its degradation and/or loss during acceleration or subsequent transport does not constitute a serious problem in view of driving a free electron laser. 
Depending on extent of beam loading, the peak current longitudinal phase space may result distorted either by accelerating field curvature (negligible beam loading) or by its local deformation (dominant beam loading). It is not possible to estimate \emph{a priori} which of those effects will dominate and need to be checked by simulations.

Another non ideal feature whose effect cannot be predicted by scalings is the unavoidable presence of plasma ramps at the capillary tips, due to gas leakage. It has been shown how these features can be fruitfully exploited to ease matching into/from plasma \cite{Toma2016}; however, this would require a lengthy optimization both of transport and of capillary engineering so, in this work, we will only assess the maximum acceptable ramp length for avoiding an excessive beam degradation, assuming the ramps have an exponential behavior.

\subsection{Laser parameters and guiding}

The main methods available nowadays for driving high power lasers are plasma transverse tapering (plasma channel) and hollow waveguides \cite{Cros2002}. Both methods have similar performances, as long as the conditions $k_l \sigma_l \gg 1$ and $k_p \sigma_l \gg 1$ are met ($k_{l,p}$ being the laser and plasma wavenumber, respectively, and $\sigma_l$ the laser rms transverse size). Of the two conditions, the latter may result to be more problematic to verify at given laser parameters. Failing to meet it, results to cause more problems for the hollow waveguide than for the plasma channel, since the plasma wave may be deformed in a not easily predictable (nor simulated) way due to the interaction with inner capillary walls. On the other hand, a hollow waveguide does not require a pre-ionized and tapered plasma, like the plasma waveguide; however, ionization and tapering can be quite easily attained either by the flow of an electric current  (discharge capillary) \cite{Zigl1996} or by an ad hoc ionizing laser pulse \cite{Durf1993}.

We stick to minimal requirements for laser energy, namely an energy of \SI{6}{\joule} on target. This is motivated by the necessity to preserve most of the laser pulse intensity for a subsequent staging up to \GeV{5}. Setting $c\tau \approx \lambda_p$ (with $\tau$ the FWHM laser pulse length) requires $\sigma_l \approx$ \um{20} and the laser resonant Lorentz factor \cite{Esar2009} is $\gamma_g = $ \num{68}. For avoiding excessive slippage and ease of operation we increase laser size to \um{35}, so that $\gamma_g =$ \num{87}. This choice sets $\tau \approx$ \SIrange{100}{110}{\femto\second}. In both scenarios $k_p \sigma_l$ is of order \num{1}, so a plasma channel must be considered. 
In Table \ref{LWFA_module} we summarize the choices made for the LWFA module.
\par

\begin{table}
\centering
\begin{tabulary}{0.95\textwidth}{||C|C|C|C|C||} 
\hline
\hline
\multicolumn{5}{||c||}{Laser parameters} \\
\hline
\hline
E [\SI{}{\joule}] & $\tau$ [\fs{}] & $\sigma_{tr}$ [\um{}] & $Z_r$ [\mm{}] & $a_0$ \\
\hline
6 & 110 & 35 & 4.8 & 1 \\
\hline
\hline
\multicolumn{5}{||c||}{Plasma parameters} \\
\hline
\hline
\multicolumn{2}{||c|}{$n_0$ [\cmmc{}]} & L [\cm{}] & \multicolumn{2}{|c||}{$R_{in}$ [\um{}]} \\
\hline
\multicolumn{2}{||c|}{\num{e17}} & 6 & \multicolumn{2}{|c||}{$\gtrsim$ 350} \\
\hline
\hline
\end{tabulary}
\caption{Laser and plasma module parameters} \label{LWFA_module}
\end{table}

\subsection{Matching line and undulator}
Matching to the undulator modules is enforced by a \SI{8}{\meter} long matching section composed by 9 electromagnetic quadrupoles \SI{0.2}{\meter} long and with a bore \cm{10} wide \cite{RosC2017}. The position and focusing gradients for those quads have been found using the genetic optimizer GIOTTO \cite{giotto} and the tracking code ASTRA \cite{astra}.

The state of the art for undulators places the shortest period actually demonstrated in a FEL experiments at $\lambda_{u}=1.4$ cm \cite{KYMA} with $a_{w}$=0.7. This means that \GeV{1} electron energy allows to reach radiation wavelengths of the order of \SI{3}{\nano\meter}. based on Ming Xie formulas \cite{mingxie}, assuming an emittance of \SI{0.5}{\milli\meter\milli\radian}, we exceed \num{e12} photons per shot in about \SI{20}{\meter} of undulator, provided that the factor $I/\sigma_{x}^{2}$ is larger than \SI{1.5}{\ampere\per\square\micro\meter} and the relative energy spread of the beam is maintained below \num{e-3} ($I$ being peak current and $\sigma_{x}$ electron transverse dimension). For a reasonable matched transverse dimension of the electron beam at the undulator entrance of about \SIrange{30}{40}{\micro\meter}, the electron current should be of the order \SI{2}{\kilo\ampere} or larger.

\section{Simulations results} \label{simulations}

\subsection{Plasma simulation settings}  
We performed simulations for the LWFA module by using the code QFluid \cite{Toma2016}, simulating propagation and transport of the input bunch in a \cm{6} long, flat top longitudinal plasma profile, either with and without input and output ramps. For ramped settings, ramps are assumed to have exponential profiles with a varying characteristic length $l_r$, namely $l_r = 1/2 \lambda_\beta$, $l_r=\lambda_\beta$ and $l_r=2\lambda_\beta$, where  $\lambda_\beta \approx$ \mm{5}  is the beam betatron wavelength in the flat top region.
In all simulations we used a spatial sampling of $\lambda_p/200$ both in the longitudinal and transverse directions, while the time step is $dt =$ \fs{1.5}. The witness bunch is sampled with \num{2e5} macro-particles.

\subsection{Simulation results: ideal plasma target}
Optimization of the ideal setting requires to find both the correct beam matched size $\sigma_m$ and injection phase $\phi_0 = k_p z_0$, with $z_0$ the initial distance between laser and beam centers of mass. The laser driver is focused at nominal matched size at the beginning of the plasma plateau at density $n_0$. Since the input beam is slightly asymmetric in $x$ and $y$ directions, we start by symmetrizing its Twiss parameters at constant emittances. Evaluating the matching condition for the equivalent fully symmetric beam (i.e. using an emittance given by the square mean root of $x,y$ values) returns a matched size of \um{0.5}: we use it as a first estimation of the correct value. Starting from there, we change both $\beta_T$, the beam injection Twiss beta function, and $\phi_0$ to find the most performing configuration; best performances are selected based on slice properties. For the best setup, emittance and energy spread slice analysis is depicted in Figure \ref{slices} (solid lines) together with the beam current profile. The final beam energy is \MeV{1065}, demonstrating an average accelerating field in excess of \GVm{9}, perfectly in line with what expected.

\subsection{Plasma simulation results: plasma target with ramps}

In simulations for the ramped settings, the driving laser pulse (de)focuses along the ramps. This means its starting (final) spot size is larger than nominal matched value and the resulting driven plasma wave is linear. Since QFluid can not operate at zero background density, we chose to start with an initial density of $n_i =$ \cmmc{e14}. 
In low density plasmas, beam loading is usually dominating, so that no sound estimate for the matched beam spotsize exists. Again, we scan initial electron beam size and injection phase for best performances. We also allow for a non zero value of the $\alpha_T$ Twiss parameter, since it has been reported \cite{Toma2016, Dorn2015} that a converging beam envelope at the ramp beginning can improve quality preservation. We notice that the witness must be injected closer to the driver with respect to the ideal setting: this is due to the fact that in the ramps, due to low plasma density and larger driver spot size, the laser group velocity is larger than in the plateau.
\begin{figure}[!h]
\begin{center} 
    \includegraphics[width=.45\textwidth]{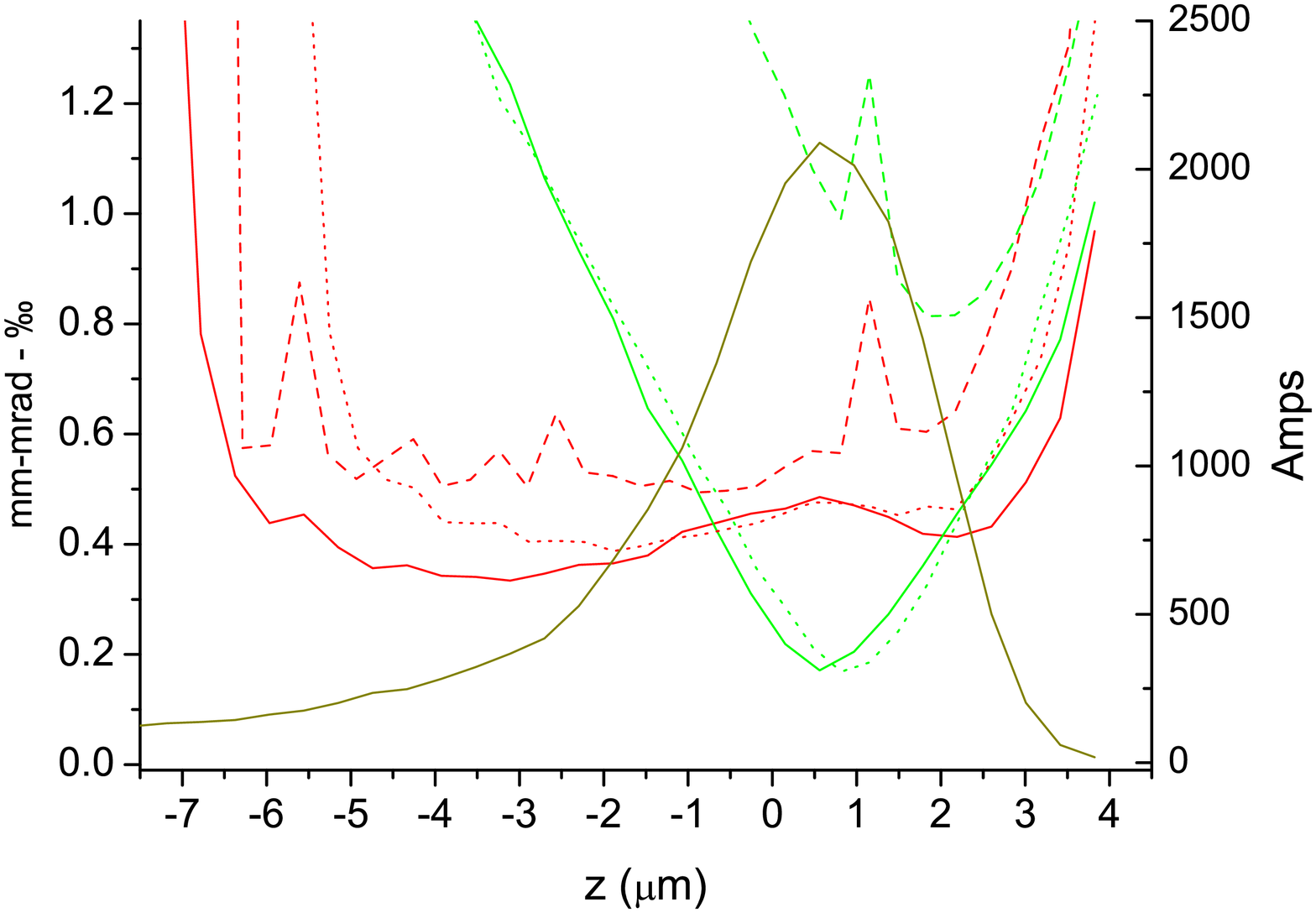}
    \caption{Slice normalized emittance (red) and rms energy spread (green) for three different plasma settings: ideal (solid lines), $l_r=1/2\lambda_\beta$ (dotted lines) and $l_r=\lambda_\beta$ dashed lines. The curren profile is also reported (dark yellow line).}
    \label{slices}
\end{center}
\end{figure}

We also performed acceleration employing longer characteristic ramp lengths, namely $l_r = \lambda_\beta$ and $l_r = 2\lambda_\beta$. Those settings are still under investigation and no optimized transport and injection conditions have been found yet. Still, we managed to obtain a satisfactory acceleration for the $l_r = \lambda_\beta$ case while did not succeed when $l_r = 2\lambda_\beta$ where the bunch is almost completely lost in plasma. It is worth noting, however, that characteristic ramp lengths in excess of few \mm{} are not likely to represent a realistic setting.  

In Figure \ref{slices}, we compare the slice analysis for ideal (solid lines) and ramped ($l_r = \lambda_\beta$ dotted lines, $l_r = \lambda_\beta$ dashed lines) settings; those values are calculated without any charge cut and show that no significant variations between the first two settings in the current peak region. As for the longer ramp scenario, the accelerated beam, whose injection phase still needs optimization, suffers from a slight but still tolerable (from FEL point of view) slice emittance and energy spread degradation. Projected beam parameters are reported in Table \ref{after} where final emittance is calculated after a $5\sigma$ cut in transverse phase spaces and $\Delta E$ is spectrum FWHM.

\begin{table}[!h]
\centering
\begin{tabular}{|c|c|c|c|}
\hline 
Quantity &  Ideal & $l_r=1/2\lambda_\beta$ & $l_r=\lambda_\beta$
\tabularnewline
\hline 
\hline 
$\beta_T$ (\mm{}) & 0.75 & 20.00 & 24.57
\tabularnewline
\hline 
$\phi_0$ & 0.465 & 0.410 & 0.410 
\tabularnewline
\hline 
$\alpha_T$ & 0 & 1 & 0 
\tabularnewline
\hline 
E (\MeV{}) & 1069 & 1048 & 1041 
\tabularnewline
\hline 
$\varepsilon_n$ (mm-mrad) & 0.43 & 0.47 & 0.47 
\tabularnewline
\hline 
$\Delta E/E$ (\%) & 0.07 & 0.09 & 1.9
\tabularnewline
\hline
$Q$ (\pC{}) & 28.2 & 28.2 & 26.7 
\tabularnewline
\hline 
\end{tabular}

\caption{\label{after}Beam parameters at injection and after acceleration.}

\end{table}

\section{FEL radiation}

\begin{table}[!h]
\centering
\begin{tabular}{|c|c|c|c|}
\hline 
Quantity  & a1 & b & a2\tabularnewline
\hline 
\hline 
$\lambda_r$  (\SI{}{\nano\meter})  & 2.68 & 2.8 & 2.7 \tabularnewline
\hline 
$\rho_{1d/3d}$  $\left(\times\num{e-3}\right)$ & 2-1.66 & 2-1.68 & 2-1.66\tabularnewline
\hline 
$\lambda_u$  (\SI{}{\meter}) & 12-25 & 11-20 & 22\tabularnewline
\hline 
Energy ($\SI{}{\micro\joule}$) & 12-25 & 21-40 & 63\tabularnewline
\hline 
Ph/pulse  $\left(\times\num{e11}\right)$ & 1.6-3.4 & 2.4-4.3 & 8.5\tabularnewline
\hline 
Bandwidth   & 1.5-2.3 & 2.5-4 & 3\tabularnewline
\hline 
\end{tabular}

\caption{\label{tab:FEL-parameters}FEL parameters.}
 
\end{table}

The FEL simulations have been all performed by means of the code GENESIS 1.3 \cite{genesis} in time-dependent mode. We directly uploaded beams phase spaces in the code, using the best possible precision options. 

Radiation characteristics are summarized in Table \ref{tab:FEL-parameters}. Using $\lambda_{u}$ = \cm{1.5} and $a_{w}$ = 0.8, the resonant wavelength is $\lambda_{r}$ = \nm{2.7}. Matching to the undulator has been done for the best beam slice, both both ideal and $l_r = 1/2 \lambda_\beta$ settings, and led to values of the rms $x$-$y$ dimensions of about \um{33} and \um{22}, respectively.  The 1-D FEL parameter turns out to be $\rho$=2 \num{e-3}, while the 3-D one is $\rho_{3D}$=\num{1.66e-3}, for a gain lengths of about \SI{0.4}{\meter}. Cooperation length is about \um{0.5} and $2\pi L_{c}$ = \um{3.1}. This corresponds, once saturation is reached, to a single spike operation ($L_{b}<2\pi L_{c}$) \cite{ssvilla,ssrosen}. 

\begin{figure}[!h]
\centering
\includegraphics[width=1\columnwidth]{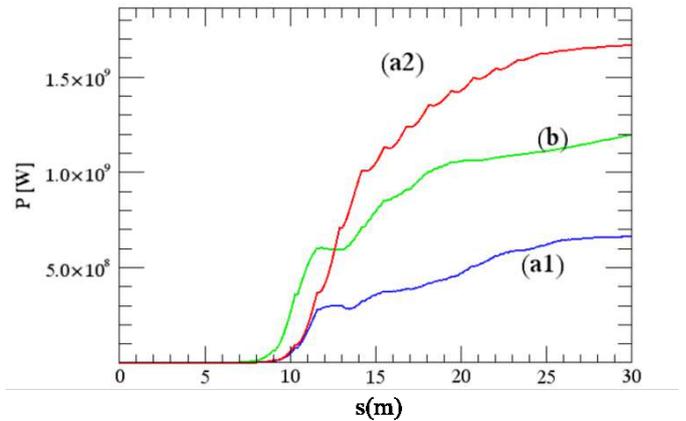}
\caption{\label{fig:Growth-of-the}Growth of the radiation in the undulator. Ideal setting with constant (a1) and tapered (a2) undulator field and setting with $l_r=1/2 \lambda_\beta$ (b). }
\end{figure}

Ideal setting beam has been simulated assuming either a flat undulator magnetic field (a1) or a tapered undulator field (case (a2)). Power growth vs $s$ is shown in Fig. \ref{fig:Growth-of-the}. 
In case (a1) saturation is reached in about \SI{12}{\meter} yielding \SI{12}{\micro\joule} of radiation emission; superradiance follows, doubling the energy after a total propagation in undulator of \SIrange{25}{30}{\meter}. A slight tapering of -2\%/m (case (a2)), optimized in slope and starting at \SI{9}{\meter}, allows to increase the energy up to \SI{63}{\micro\joule} at \SI{23}{\meter}. Case (a1) is almost in single spike regime while, in the tapered mode, power density and spectrum present a few oscillations. 

Case (b) represents the $l_r = 1/2 \lambda_\beta$ setting. The electron beam emission results to be improved with respect to the ideal setting, because the presence of the ramps gives further degrees of freedom for beam shaping. Also in this case, a first saturation at \SI{11}{\meter} gives \SI{24}{\micro\joule}, followed by a further growth to \SI{43}{\micro\joule}, where the total number of photons reaches \num{6 e11} per shot.

\section{Conclusions} \label{conclusions}
In this paper we performed simulations for a laser driven plasma booster enabling to double the energy of a \MeV{500} incoming high brightness electron bunch preserving beam quality. Different plasma configurations have been studied, showing how the presence of exponential ramps at the plasma target ends does not impact significantly on plasma booster performances if their length does not exceed the beam betatron wavelength. Electrons are then matched to a short period undulator and radiation production by FEL instability at \nm{2.7} is demonstrated by GENESIS simulations. Again, different undulator configurations are checked each of them producing an $O(\num{e11})$ photons per shot.

\section*{Acknowledgements}
This work has been partially supported by the European Union's Horizon 2020 research and innovation program under grant agreement No. 653782

\bibliographystyle{elsarticle-num}



\end{document}